\documentclass[preprint2]{aastex}

\slugcomment{}

\shorttitle{The Sub-Luminal Parsec-scale Jet of Mkn 501}
\shortauthors{Edwards \& Piner}

\begin{document}

\title{The Sub-Luminal Parsec-scale Jet of Mkn 501}

\author{P. G. Edwards}
\affil{Institute of Space and Astronautical Science, Yoshinodai, Sagamihara,
       Kanagawa 229-8510, Japan}
\email{pge@vsop.isas.ac.jp}
\and
\author{B. G. Piner}
\affil{Department of Physics \& Astronomy, Whittier College, 
       13406 E. Philadelphia St., Whittier CA 90608, U.S.A.}
\email{gpiner@mail.whittier.edu}

\begin{abstract}
We have combined Very Long Baseline Interferometry (VLBI) data from
several programs in order to resolve differences in reported
parsec-scale jet speeds for the TeV gamma-ray source Mkn\,501.  Data
from the VLBA 2cm survey, and 8 and 15\,GHz data from the Radio
Reference Frame Image Database, have been combined with data from a
5\,GHz VSOP Space VLBI observation to determine the apparent motions
of jet components in this source.  The combined data set consists of
12 observations between April 1995 and July 1999. Four jet components
are detected at most epochs, all of which are clearly sub-luminal
(i.e. with apparent speeds $< c$), and two of which appear stationary.
The established TeV gamma-ray sources Mkn\,501 and Mkn\,421 thus both
have sub-luminal parsec-scale jets, in contrast to the apparently
super-luminal jets of the majority of $>$100~MeV sources detected by
EGRET. No new VLBI component has emerged from the core 
following the extended TeV high-state in 1997, again in contrast to
the general behavior of $>$100~MeV gamma-ray sources.
\end{abstract}


\section{Introduction}

The BL~Lac object Mkn\,501 (1652+398, J1653+3945, DA426) has been
well-studied at radio wavelengths \citep[e.g.][]{muf84,gab92}, 
but interest in the source was rejuvenated by the discovery of TeV
gamma-ray emission \citep{qui96}.  A prolonged high-state at TeV
energies in 1997 included activity on timescales of several hours,
implying the TeV gamma-rays originate in a relatively compact area
\citep[see, e.g., the review of][]{cat99}. The detection of
correlated X-ray and TeV gamma-ray variability in the other
well-studied TeV source, Mkn\,421, is strong evidence in favor of the
X-ray emission being the high end of the synchrotron component of the
spectral energy distribution, with the TeV emission arising from
inverse Compton scattering of photons by the synchrotron emitting
electrons, and such a model is also widely accepted for Mkn\,501.

On the parsec scale, VLBI observations have revealed that a jet
emerges from the core at a position angle of $\sim$180$^\circ$, and
bends by $\sim$90$^\circ$ within the first $\sim$2\,milli-arcseconds
(mas).  The jet extends to the east until $\sim$20\,mas from the core,
when it bends further, finally reaching the position angle of
$\sim$45$^\circ$ seen on the kilo-parsec scale \citep{con95,gio99}.
The parsec-scale jet of Mkn\,501 is one-sided, and it is assumed that
this jet is relativistically doppler boosted, while the counter-jet is
doppler de-boosted to such an extent it is invisible at the current
sensitivity of VLBI observations. The jets are believed to originate
in the accretion disk surrounding a central super-massive black hole,
which for Mkn\,501 has been suggested to have a mass of
10$^{8.93\pm0.21}$\,M$_\odot$ (Falomo, Kotilainen \& Treves 2002).

A number of different motions have been reported for components in the
Mkn\,501 jet, ranging from 0.27$\pm$0.02\,mas\,yr$^{-1}$ \citep{gab94}
to 2.4\,mas\,yr$^{-1}$ \citep{gio99}.  These studies have relied on
only a small number of epochs, typically less than four, and, in
hindsight, have probably under-estimated the errors in locating jet
components. As illustrated by \citet{pin99} for Mkn\,421,
reliable determination of component motions generally requires larger,
multi-epoch data sets.

Here we study the parsec-scale jet of Mkn\,501 from twelve VLBI
observations spanning 4.28 years, at frequencies of 5, 8 and 15\,GHz.
Mkn\,501 lies at a redshift of 0.034 \citep{wil74} which, for the
value of H$_0 = 65$\,km\,s$^{-1}$\,Mpc$^{-1}$ adopted throughout his
paper, corresponds to a distance of 155\,Mpc.  At this distance, an
angular separation of 1\,mas corresponds to a projected linear
distance of 0.72\,pc.

\begin{table*}
\caption{Observations of Mkn 501}
\begin{center}
\begin{tabular}{l c r r r c l} \tableline \tableline
                      & Frequency &  \multicolumn{3}{c}{Synthesized Beam\tablenotemark{a} } & Image r.m.s.\ Noise & \\ 
 \multicolumn{1}{c}{Date} &  (GHz) &  (mas) & (mas) & ($^\circ$) & ($\mu$Jy beam$^{-1}$)  & \multicolumn{1}{c}{Program}  \\ \tableline
1995 Apr 8  & 15 & 0.96 & 0.52 &  $-$5.2 & ~390 & 2 cm survey \\
1995 Apr 12 & ~8 & 1.33 & 1.26 &    18.4 & ~675 & RRFID       \\
1995 Oct 17 & 15 & 1.24 & 0.59 &  $-$1.6 & 1195 & RRFID       \\
1995 Dec 15 & 15 & 0.96 & 0.54 &  $-$7.1 & ~320 & 2 cm survey \\
1996 Apr 23 & 15 & 0.85 & 0.66 &     2.2 & ~565 & RRFID       \\
1996 Apr 24 & ~8 & 1.89 & 1.22 &    19.9 & ~445 & RRFID       \\
1996 Jul 10 & 15 & 0.96 & 0.53 &  $-$1.4 & ~275 & 2 cm survey \\
1997 Mar 13 & 15 & 0.95 & 0.54 &  $-$8.8 & ~205 & 2 cm survey \\
1998 Apr 7  & ~5 & 0.58 & 0.23 &    21.9 & ~570 & VSOP        \\
1998 Jun 24 & ~8 & 0.94 & 0.76 & $-$19.7 & ~515 & RRFID     \\
1998 Oct 30 & 15 & 0.90 & 0.53 &     2.9 & ~300 & 2 cm survey \\
1999 Jul 19 & 15 & 1.26 & 0.61 &  $-$8.7 & ~290 & 2 cm survey \\ \tableline
\end{tabular}
\end{center}
\tablenotetext{a}{Numbers given for the beam are the FWHMs of the major
and minor axes in mas, and the position angle of the major axis in degrees.}
\end{table*}

\section{VLBI observations}

We have compiled data from three programs for this study: the Very
Long Baseline Array (VLBA) 2\,cm survey \citet[][see also
http://www.cv.nrao.edu/2cmsurvey]{kel98}, the Radio Reference Frame
Image Database \citep[][see also
http://rorf.usno.navy.mil/rrfid.shtml]{fey97}, and a single VLBI Space
Observatory Programme \citep[VSOP;][]{hir98,hir00} observation.  The
data used in this study are summarized in Table~1.

The VLBA 2\,cm survey is being undertaken at multiple epochs to study
the properties and evolution of over 100 active galactic nuclei.  The
Mkn\,501 observations consisted of eight scans at one hour intervals
of typically 5~minutes duration.  Data were recorded with a bandwidth
of 64\,MHz using 1~bit samples and left-circular polarization.  An
image derived from the March 1997 data used in this paper was
published by \citet{kel98}.  The full ten-station array was used at
all epochs except the last, for which the Saint Croix telescope was
unavailable.

The Radio Reference Frame Image Database (RRFID) of the U.S.\ Naval
Observatory (USNO) is a program to regularly image the radio sources
used for precise astrometry.  The VLBA was used at all epochs,
although the October 1995 observation was made without the Mauna Kea
and North Liberty telescopes.  The June 1998 observation was made with
the addition of the Fairbanks 26\,m (Alaska), Green Bank 20\,m (West
Virginia), Kokee Park 20\,m (Hawaii), Medicina 32\,m (Italy), Ny
Alesund 20\,m (Norway), Onsala 20\,m (Sweden), and Westford 18\,m
(Massachusetts) telescopes.  Typically, four scans of $\sim$3 minutes
were made, with bandwidths of 16\,MHz for the first two epochs,
32\,MHz for the 8\,GHz observations of the last two epochs, and
64\,MHz for the 15\,GHz observation in April 1996. Right circular
polarization is recorded for all RRFID observations.  An image from
the 1995 April 12 epoch at 8.4\,GHz was published by \citet{fey97}.

The 5\,GHz VSOP observation, in April 1998, was made over a 13 hour
period with the HALCA satellite, the VLBA and the Effelsberg 100\,m
(Germany) telescope.  Interferometric fringes to the satellite were
detected from tracking passes totaling 7\,hours.  In the standard VSOP
observing mode, 32\,MHz of two-bit sampled, left circular polarization
data is recorded.  Although made at the lowest frequency considered
here, the long baselines to the orbiting telescope result in the
synthesized beam-size for this observation being the smallest of these
data.  The VSOP data considered here were combined with a 1.6\,GHz
VSOP observation one day later to derive a spectral index map of the
source \citep{edw00a}.

\section{Analysis}

The data were fringe-fit in AIPS and we have imaged all data ourselves
using the Difmap package.  Beams were calculated using natural
weighting (uvweight=0,$-$1 in Difmap), with the exception of the VSOP
observation for which uniform weighting (uvweight=2,0 in Difmap) is more
appropriate \citep{hir00}.  
An image from the 15\,GHz RRFID observation
in April 1996 is shown in Figure~1.

Model-fitting of the images was carried out in Difmap.  As inspection
of Figure~1 reveals, at most epochs four jet components were required
in addition to the core to provide a good representation of the data.
We have labeled these C1 through C4, with C1 being the component
farthest from the core.  Circular gaussian components were fitted at
all epochs.  Full details of the model-fits 
are given in Table~2.
Reduced $\chi^2$ values for the fits are not given in the
Table, as they are dependent on the way the data from the
different programs was reduced. 
Thus, while model fits represent a minimum in $\chi^2$ for the given
number of components, the comparison of $\chi^2$ values between data from
different programs is potentially misleading 
\citep[see also][]{pin99}.

\begin{table*}
\caption{Gaussian Model Fits to Source Components}
\label{mfittab}
\begin{center}
{\small \begin{tabular}{l c c r r r r} \tableline \tableline
& Frequency & Component
& \multicolumn{1}{c}{$S$\tablenotemark{a}} & \multicolumn{1}{c}{$r$\tablenotemark{b}} &
\multicolumn{1}{c}{PA\tablenotemark{b}} &
\multicolumn{1}{c}{$a$\tablenotemark{c}} \\
\multicolumn{1}{c}{Epoch} & (GHz) & ID
& \multicolumn{1}{c}{(mJy)} & \multicolumn{1}{c}{(mas)} &
\multicolumn{1}{c}{(deg)}
& \multicolumn{1}{c}{(mas)} \\ \tableline
1995 Apr 8  & 15 & Core & 489 & ...  & ...   & 0.17 \\
            &    & C4   & 114 & 0.75 & 172.1 & 0.63 \\
            &    & C3   & 84  & 2.32 & 147.7 & 1.25 \\
            &    & C2   & 57  & 4.01 & 133.1 & 1.95 \\
            &    & C1   & 62  & 7.45 & 113.0 & 2.58 \\ \tableline
\end{tabular}}
\end{center}
\tablenotetext{a}{Flux density in milliJanskys.}
\tablenotetext{b}{$r$ and PA are the polar coordinates of the
center of the Gaussian relative to the core.
Position Angle is measured from north through east.}
\tablenotetext{c}{$a$ is the FWHM of the Gaussian.}
\tablenotetext{}{
[The complete version of this table is in the electronic edition of
the Journal.  The printed edition contains only a sample.]
}
\end{table*}

The positions of all model-fit jet components are plotted as a
function of time in Figure~2. In order to determine the uncertainty in
the component location for the purposes of determining component
motions, we have projected the beam major-axis onto the line joining
the component and core and then taken a fraction of this projected
length as the error in position.  For the extended, outermost
component, C1, we conservatively adopted half a projected beamwidth
for the uncertainty.  For C2 and C3 we used one quarter of the
projected beamwidth, and for the innermost component, C4, we used
one-eighth of the beamwidth.  Motions were determined by weighted
linear fits to the data.  As shown in Figure~2, both C1 and C4 show
little evidence of motion over the 4.28\,year period.  Both C2 and C3
show clear evidence of motion, with an apparent component speeds
for C2 of 0.6$\pm$0.1\,$c$, and for C3 of 0.3$\pm$0.1\,$c$.  
The apparent speeds along the jet, as opposed to radial separations,
are, within errors, the same.
These speeds supersede the preliminary values
reported in \citet{pin02}.

\section{Discussion}

Our component locations agree well with those reported from a
contemporaneous 5\,GHz observation made in June 1996 as part of the
VLBA Pre-launch Survey \citep{fom00}.  We can also extrapolate our
derived motions and compare them with model-fits to the 5\,GHz
observations at epochs 1987.4 \citep{gab92}, and 1989.3 \citep{gab94}.
The extrapolated motion of C3 is consistent with the positions of the
K2 of \citet{gab92} and \citet{gab94}, assuming uncertainties $\sim$3
times larger than the $\pm$0.1\,mas adopted by these workers.  The K1
of \citet{gab92} and \citet{gab94}, with similarly increased
uncertainties, is consistent with the extrapolated motion of C2,
particularly if the speed of C2 lies at the lower end of the range
determined in \S3 (assuming a constant motion).  These identifications
were qualitatively suggested by \citet{edw00b}, but are quantitatively
borne out by the fuller analysis presented here.

The speeds of C4 and C1 are formally consistent with zero, i.e.\ they
appear to be stationary components.  Before considering this further,
we reconcile this result for C4 with the observations of \cite{mar99},
who reported the detection at 22\,GHz of a resolved component between
0.5 and 1\,mas from the core, with an apparent motion derived from
three epochs between April and August 1997 of
0.96$\pm$0.1\,mas\,yr$^{-1}$, corresponding to 2.3$\pm$0.2\,$c$.  This
location is consistent with our C4, however we do not see such rapid
motion over the four year period. The motion reported by Marscher
corresponds to 0.27\,mas in the 0.29 years the observations
spanned. In our data, C4 ranges between 0.59\,mas from the core (1998
April~7) and 0.83\,mas from the core (1999 July~19), a range of
0.21\,mas, similar in magnitude to that of \cite{mar99}.  Any attempt
at further interpretation is complicated by the fact that there are
likely to be frequency-dependent offsets in the separation of
components from the core \citep[see, e.g.,][]{lob98} in our data,
which would be most important for C4.

Stationary components have been reported for a number of sources in
the past, with a detailed study being made as part of the multi-epoch
monitoring program of \citet{jor01a}.  The monitoring revealed that
the super-luminal speeds detected for these of EGRET-detected blazars
were much faster than for the general population of bright compact
radio sources, however evidence was also found for at least one
stationary component in 27 of the 42 sources \citep{jor01a}.  Jorstad
et al.\ suggested that the stationary components within several
parsecs of the core were associated with standing recollimation shocks
caused by pressure imbalances at the boundary between the jet and the
surrounding medium. In contrast, the stationary components further
from the core tended to be associated with bends in the parsec-scale
jet.  There is support for this scenario in our data. C4 is located at
a projected distance of $\sim$0.5\,pc from the core, and is quite
plausibly associated with a recollimation shock.  C1, on the other
hand, is an extended component, which 1.6\,GHz VLBI imaging has
revealed is associated with a significant change in the jet from a
bright ``spine'' to a limb-brightened morphology \citep{gio99}.

If we assume our fastest observed pattern speed (0.6$c$ for C2)
reflects the bulk apparent speed of the jet, then we can solve for the
intrinsic speed and angle to the line-of-sight, provided we also have
an estimate of the Doppler beaming factor.  A Doppler factor
$\delta\sim 10$ is inferred from the TeV observations of this source
(e.g., Tavecchio, Maraschi, \& Ghisellini 1998).  The VSOP
observations yield our best measurement of the radio-core brightness
temperature, 4$\times10^{11}$\,K.  This is consistent with a Doppler
factor of $\sim$10 if the source is in equipartition \citep{rea94},
but it is also consistent with lower Doppler factors if equipartition
is violated \citep{kel02}.  If we accept the values of 0.6$c$ and
10 for the apparent bulk speed and Doppler factor, then the Lorentz
factor of the Mkn\,501 jet is $\gamma=5$ ($v=0.98c$) and its angle to
the line-of-sight is $\theta=0.7\arcdeg$.  Such a small angle to the
line-of-sight may be expected of a $\gamma$-ray blazar, although
subluminal apparent speeds are in general not expected 
\citep[see the Monte Carlo simulations of][]{lis99}. 

Alternative kinematics that do not place such tight constraints on the
angle to the line-of-sight assign the Doppler factor measurement to
the TeV-emitting region (on the light-day size scale) and the apparent
bulk speed to the VLBI jet (on the light-year size scale), and allow a
change in the bulk Lorentz factor or angle to the line-of-sight in the
intermediate region.  If the jet in the TeV-emitting region has, e.g.,
$\theta=5\arcdeg$ and $\gamma=7$ (enforcing $\delta=10$), then a
decrease in the Lorentz factor to $\gamma=2$ would reproduce the
observed apparent speed in the VLBI jet.  Such a deceleration of
electron-positron jets close to the core is proposed by \citet{mar99} 
for the TeV blazars.  A change in angle to the line-of-sight,
perhaps accompanying the large bend in the jet seen $\sim$2\,mas from
the core, cannot by itself reproduce the observed values; a jet with
$\delta=10$ has a minimum Lorentz factor of 5, and a jet with
$\gamma=5$ can only have an apparent speed of 0.6$c$ in the
large-angle solution for $\theta>90\arcdeg$.  Any set of kinematic
parameters must also be constrained by the one-sided appearance of the
source; the example above with $\theta=5\arcdeg$ and $\gamma=2$ would
have a jet-to-counterjet brightness ratio greater than $\sim$200, somewhat
higher than the limit that can be placed from our observations.

Similar values for the Doppler factor and apparent jet speed apply to
the other well-studied TeV blazar, Mkn 421 \citep{pin99}.  From
these two sources, it appears that TeV blazars as a class may either
have very small angles to the line-of-sight ($\theta<1\arcdeg$), or
may decelerate significantly between the TeV-emitting region and the
parsec scale.

It is notable that no new component has emerged from the core after
the prolonged TeV high-state in 1997.
A component with a speed similar
to that of C2 or C3 would now be $\sim$0.5\,mas from the core and
would have been detected at the latter epochs.  This suggests that
events which give rise to extended TeV (and associated X-ray) activity
are different in nature to those which result in the production of new
VLBI components \citep[see also][]{mar99}. 
Mkn\,421 and Mkn\,501, for which the inverse-Compton
component of the spectral energy distribution (SED) peaks at TeV
energies, have sub-luminal component speeds and apparently no new
component emerging after epochs of TeV activity.  In contrast, sources
with the inverse-Compton component of the SED peaking at GeV energies
tend to have the emergence of new, super-luminal, VLBI components
associated with GeV flaring states \citep{jor01b}.  The detection of
more TeV gamma-ray sources by the next generation of air Cerenkov
telescopes will enable these apparent trends to be investigated more
quantitatively.

\acknowledgments

This research has made use of the United States Naval Observatory
Radio Reference Frame Image Database (RRFID), and the NASA/IPAC
Extragalactic Database (NED) which is operated by the Jet Propulsion
Laboratory, California Institute of Technology, under contract with
the National Aeronautics and Space Administration.  
Ken Kellermann and Alan Fey are particularly thanked for the provision
of fringe-fit data from the VLBA 2cm survey and RRFID observations, 
respectively.
The National Radio Astronomy Observatory is a facility of the National
Science Foundation operated under cooperative agreement by Associated
Universities, Inc.
We gratefully acknowledge the VSOP Project, which is led by the 
Institute of Space and Astronautical Science in cooperation
with many organizations and radio telescopes around the world.
BGP acknowledges support from Whittier College's Newsom Endowment.

\clearpage 

\begin{figure}
\epsscale{0.8}
\plotone{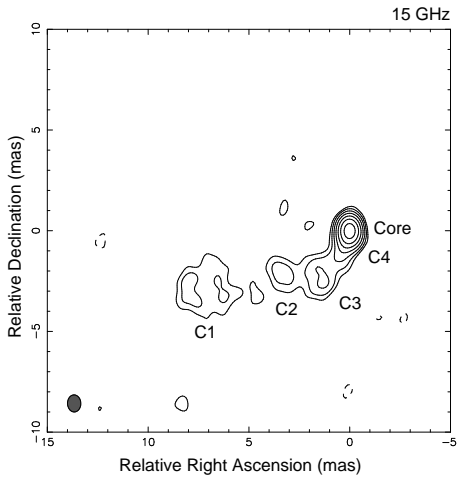}
\caption{Image of Mkn\,501 at 15\,GHz from the
RRFID observation in April 1996. The positions of the core and jet 
components are indicated (see Table~2 for details).
The beam, 0.85\,mas $\times$ 0.66\,mas (FWHM) at a position angle of 2$^\circ$,
is shown at bottom left. The contours are $-$1 (dashed), 1, 2, 4, 8, 16, 32
and 64\% of the map peak of 485~mJy/beam. \label{fig1}}
\end{figure}


\begin{figure}
\epsscale{0.8}
\plotone{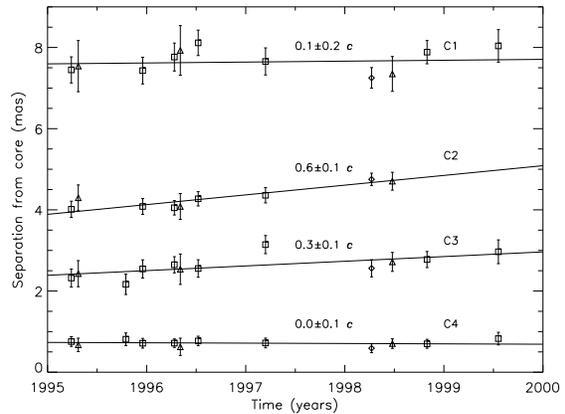}
\caption{Component positions and weighted linear fits to component motions.
The squares denote 15\,GHz observations, triangles denote 8\,GHz observations, 
and diamonds are used for the 5\,GHz VSOP observation. Note that the two
observations in April 1996 have been offset from each other in the plot 
for clarity. \label{fig2}}
\end{figure}

\end{document}